\documentclass[aps,singlecolumn,showpacs,preprintnumbers,singlecolumn]{revtex4-2}
\usepackage{float}
\usepackage{dblfloatfix}
\usepackage{fixltx2e}
\restylefloat{table}
\usepackage{graphicx}  
\usepackage{subfig}
\usepackage{subfloat}
\usepackage{multirow}
\usepackage{fancyhdr}
\usepackage{parskip}
\usepackage[T1]{fontenc}
\usepackage{dcolumn}   
\usepackage{bm}        
\usepackage{amsfonts}  
\usepackage{amsmath}   
\usepackage{amssymb}   
\usepackage{makecell}
\usepackage{caption}
\usepackage{xcolor}
\usepackage{comment}

\begin{document}

\title{Einstein's theory of Brownian motion: The non-Fickian normal diffusion}

\author{A. Bhattacharyay}
\email{a.bhattacharyay@iiserpune.ac.in}
\affiliation {\it Indian Institute of Science Education and Research, Pune, India}
\date{\today}


\begin{abstract}
{Einstein's theory of diffusion defines what is considered {\it normal diffusion}. Historically, Einstein considered homogeneous diffusion characterised by a diffusion constant in his original calculations, resulting in what is known as {\it Fickian} diffusion, which thermally relaxes concentration inhomogeneities. However, in the same derivation of Einstein, the seed of generalisation of the theory to {\it non-Fickian} diffusion in the presence of space-dependent diffusivity is hidden. This indicates that non-Fickian diffusion, which can drive concentration inhomogeneities, can also be normal diffusion. The present article discusses the origin of the {\it normal} non-Fickian diffusion in Einstein's theory of diffusion.}
\end{abstract}

\maketitle

Albert Einstein provided the physical basis of Fickian diffusion in 1905, building on his theory of Brownian motion \cite{einstein1905motion, crispin2009stochastic}. Fickian diffusion is the phenomenon of thermal transport that relaxes concentration inhomogeneities in, for example, a collection of non-interacting particles that respond to thermal fluctuations at a constant temperature everywhere in space. Fickian diffusion \cite{philibert2006one} is therefore at the core of any thermal relaxation of concentration gradients present in the spatial distribution of particles. In the presence of interparticle interactions or external fields, the competition between interactions and Fickian diffusion determines the system's steady-state concentration profile. The local Fickian diffusion current density $j_F(\bf r)$ is given by the equation
\begin{equation}
    j_F({\mathbf{r}},t) = -D\nabla C({\mathbf{r}},t),
    \label{eqn.1}
\end{equation}
where $C(\mathbf r, t)$ is the concentration of particles at position $\bf r$ at a time $t$, and $D$ is the diffusion constant or diffusivity. The diffusion constant is proportional to the ensemble-averaged squared displacement of a diffusing particle per unit time at the large time limit. This diffusive transport of particles becomes far more interesting and counterintuitive than our standard notions of uniform diffusion in the same settings as Einstein's theory of Brownian motion, when a particle's diffusivity is not constant in space, i.e., it is $D(\mathbf{r})$. In such a situation, non-Fickian diffusion arises essentially from Einstein's theory of Brownian motion, and this aspect has not received detailed attention, probably due to our long-held notion that diffusion only relaxes concentration inhomogeneities, which is actually a special case. In the present article, I will discuss the emergence of non-Fickian diffusion naturally from a straightforward generalisation of Einstein's theory of Brownian particles.

The Fickian diffusion current density as defined in eqn.(\ref {eqn.1}), when combined with the conservation of particle number through the equation of continuity, gives us the well-known diffusion equation, as derived by Einstein, for the dynamics of the concentration under Fickian diffusive transport as
\begin{equation}
    \frac{\partial C(\mathbf r)}{\partial t} = D \nabla^2 C(\mathbf r).
    \label{eqn.2}
\end{equation}
The solution of the diffusion equation (eqn.(\ref{eqn.2})), for an initial isotropic concentration at the position ${\mathbf r} = 0$ being $C(0)$ is
\begin{equation}
    C({\mathbf r},t) = \frac{C(0)}{(4\pi Dt)^{3/2}}\exp{-\frac{r^2}{4Dt}},
\end{equation}
which indicates a Gaussian spread of the initial concentration over time, with a linear-in-time, ever-growing width of the Gaussian. Historically, this phenomenon is regarded as diffusion, with its physical origin explained by Einstein's theory of Brownian motion in 1905.

It is generally assumed in standard literature that diffusion processes which differ from Einstein's theory of diffusion are {\it anomalous diffusion}. In this context, the question naturally arises: What is Einstein's theory of diffusion? Is it the equation (\ref{eqn.2}) with a constant diffusivity $D$ everywhere in space (as originally taken by Einstein), or the broader scope of the theory based on the fundamental postulates that Einstein had taken into account to formulate the theory? In this context, I underscore that Einstein's original method for deriving equation (\ref{eqn.2}) admits a straightforward generalisation to include a space-dependent diffusivity $D(x)$ (in one dimension, without loss of generality, which we will consider from now onward) while remaining consistent with the basic postulates of the theory. The resulting diffusion equation includes a non-Fickian ({\it normal diffusion}) current on top of the Fickian one mentioned above. There is no anomaly associated with the non-Fickian diffusion process that arises from Einstein's theory, as it is fully consistent with Einstein's formulation of diffusion based on Brownian motion. 

The term {\bf non-Fickian} diffusion has so far been generally used in the existing literature to particularly mean anomalous diffusion \cite{el2002non, de2005viscoelastic, vasconcellos2006statistical, petropoulos2009beyond, ferreira2015new, cherstvy2019non, matias2023model}. Even in the textbook {\it The Mathematics of Diffusion} by J. Crank \cite{crank1979mathematics}, which devotes a whole chapter to non-Fickian diffusion, it is described as primarily anomalous diffusion. For these reasons, the present author has not, so far, used the term non-Fickian diffusion extensively in his earlier papers, although non-Fickian diffusion can also be perfectly normal \cite{bhattacharyay2019equilibrium,bhattacharyay2020generalization,bhattacharyay2025brownian}. However, the present author thinks it is time to define the term non-Fickian diffusion in its most appropriate sense relative to Fickian diffusion, both of which are normal, i.e., consistent with Einstein's theory of Brownian motion. It is the purpose of the present article to establish non-Fickian diffusion as {\it normal} diffusion as much as the Fickian diffusion itself.

The Fickian and non-Fickian (as described below) diffusion processes are consequences of the same theory of Brownian motion introduced by Albert Einstein, without altering any of the fundamental assumptions he set forth. While Fickian diffusion relaxes concentration inhomogeneities because a constant diffusivity corresponds to an underlying homogeneous space, non-Fickian diffusion drives them because the diffusivity is space-dependent, reflecting the inhomogeneity of the underlying space in which the broken symmetry must show up in the transport process. However, at the mesoscopic scales where Einstein's Brownian motion drives the process, both diffusion processes are manifestations of the same physics, with even the diffusivity defined exactly in the same way. It will be of further interest to examine the historical reasons for treating only Fickian diffusion as {\it normal}, while Einstein's theory effectively lays the foundation for diffusion in general, including both Fickian and non-Fickian cases.


Let us follow the essential details of Einstein's theory of Brownian motion as presented in the textbook on Stochastic Methods by Crispin Gardiner \cite{crispin2009stochastic} in line with the original paper by Albert Einstein. It will be followed by a straightforward generalisation of Einstein's method to define space- or coordinate-dependent diffusivity and diffusion current densities, which will show that {\it normal} diffusion not only relaxes concentration inhomogeneities but can also drive them.

Einstein's theory of Brownian motion is based on two postulates, which could be summarised as:
\begin{itemize}
    \item The Brownian motion is a slower transport process, being an average motion due to exceedingly large collisions the Brownian particle suffers in the intervening microscopic times with the molecules of fluid.
    \item The stochastic displacements of the Brownian particle in consecutive intervals of mesoscopic times are independent owing to the local equilibrium the particle is always in within the slow time scales at which the Brownian motion occurs.
\end{itemize}
These two postulates construct the bedrock of Einstein's theory of Brownian motion, a consequence of which is the {\it normal} diffusion.

Following in Einstein's footsteps, consider the independent Brownian motion of $n$ particles in a fluid. Consider also that the position coordinate of a Brownian particle $x$ will change by the amount $\Delta$ in a time interval $\tau$. The displacement $\Delta$ follows a distribution $\phi(\Delta)$, where $dn$ particles suffer displacements between $\Delta$ and ($\Delta + d\Delta$). This means,
\begin{equation}
    dn = n\phi(\Delta)d\Delta,
    \label{eqn.2.1}
\end{equation}
where $\int_{-\infty}^{\infty}{\phi(\Delta)d\Delta}=1$ and $\phi(\Delta)=\phi(-\Delta)$ to impose the jump probability normalisation and the local isotropy of space, respectively. Note that, although the limit in the integral over jump size $\Delta$ is extended from minus to plus infinity, in all practical senses the distribution $\phi(\Delta)$ is restricted to mesoscopic scales, which are bigger than microscopic but much smaller than macroscopic \cite{bhattacharyay2025brownian}. In general, the number density of particles $f(x,t)$ is a function of position $x$ and time $t$. Now, the relation between the displacement distribution $\phi(\Delta)$ and the diffusivity (diffusion constant) comes from the following calculations.

To compute the distribution of the particles at time $t+\tau$ from that at time $t$, Einstein considered, which has become the general method in many ways in the subsequent development of stochastic theories \cite{crispin2009stochastic}, the equality
\begin{eqnarray}
    f(x,t+\tau)dx = dx\int_{-\infty}^{\infty}{f(x+\Delta)\phi(\Delta)d{\Delta}},
    \label{eqn.2.2}
\end{eqnarray}
where, obviously, $\phi(\Delta)$ is assumed independent of space, and that is what makes the diffusion a homogeneous one in space in Einstein's original calculations. Keep in mind that $x$ and $\Delta$ are independent variables, where the former specifies the position of the particle at time $t$, the latter denotes the size of the jump that the particle makes in the subsequent interval $\tau$ from the position $x$. Note that nothing prevents us from considering $\phi(\Delta)$ to be inhomogeneous on mesoscopic scales to generalise Einstein's theory of Brownian motion, as that generalisation will remain completely consistent with Einstein's postulates. However, before we embark on that generalisation, let us first have a complete look at Einstein's theory.

At this stage, upon considering the smallness of the mesoscopic time $\tau$ in comparison to the macroscopic time scale, Einstein Taylor expands $f(x, t+\tau) = f(x,t)+\tau\frac{\partial f(x,t)}{\partial t}$. Moreover, considering the displacement $\Delta$ suffered by a particle in time $\tau$ to also remain mesoscopic, the expansion of $f(x+\Delta, t)$ follows:
\begin{eqnarray}
    f(x+\Delta,t) = f(x,t) + \Delta\frac{\partial f(x,t)}{\partial x}+\frac{\Delta^2}{2!}\frac{\partial^2 f(x,t)}{\partial x^2}+ \cdot\cdot\cdot
    \label{eqn.2.3}
\end{eqnarray}
Now, inserting these expansions in eqn.(\ref{eqn.2.2}), one gets
\begin{equation}
    f(x,t)+\tau\frac{\partial f(x,t)}{\partial t} = f(x,t)\int_{-\infty}^{\infty}{\phi(\Delta)d\Delta +\frac{\partial f(x,t)}{\partial x}}\int_{-\infty}^{\infty}{\Delta\phi(\Delta)d\Delta}+\frac{\partial^2f(x,t)}{\partial x^2}\int_{-\infty}^\infty{\frac{\Delta^2}{2}\phi(\Delta)d\Delta} + \cdot\cdot\cdot ,
    \label{eqn.2.4}
\end{equation}
which, on consideration of the isotropy of $\phi(\Delta)$ and its normalisation results, at the leading order, in the diffusion equation
\begin{eqnarray}
    \frac{\partial f(x,t)}{\partial t}=D\frac{\partial^2f(x,t)}{\partial x^2},
    \label{eqn.2.5}
\end{eqnarray}
where the diffusion constant is $D=\int_{-\infty}^{\infty}{\frac{\Delta^2}{2\tau}\phi(\Delta)d\Delta}$. Note one important feature of Einstein's definition of the diffusivity in eqn.(\ref{eqn.2.5}): it is by definition local at position $x$ in time interval $\tau$. The underlying space is considered homogeneous here, given that $\phi(\Delta)$ is independent of the position coordinate. This homogeneity of space and the independence of consecutive jump events lead to an effective extension of the definition of the diffusion constant at large times, making it easier to measure experimentally. Let us check what happens when the jump distribution $\phi(\Delta)$ is a function of position. 


Considering the jump distribution function position dependent, that is $\phi(x,\Delta)$, eqn.(\ref{eqn.2.2}) should be written as 
\begin{equation}
        f(x,t+\tau)dx = dx\int_{-\infty}^{\infty}{f(x+\Delta)\phi(x+\Delta,\Delta)d{\Delta}}.
        \label{eqn.2.6}
\end{equation}
In the integrand on the right-hand side of the above equation, $f(x+\Delta)\phi(x+\Delta,\Delta)$ is the number of particles in unit length at position $x+\Delta$ making a jump of size $\Delta$ according to the local distribution $\phi(x+\Delta, \Delta)$. It will result in the need to consider the Taylor expansion of the function $\psi(x+\Delta,\Delta;t)=f(x+\Delta,t)\phi(x+\Delta,\Delta)$ as
\begin{eqnarray}
    \psi(x+\Delta,t) = \psi(x,t) + \Delta\frac{\partial \psi(x,t)}{\partial x}+\frac{\Delta^2}{2!}\frac{\partial^2 \psi(x,t)}{\partial x^2}+ \cdot\cdot\cdot\, .
    \label{eqn.2.7}
\end{eqnarray}
Now, using the eqn.(\ref{eqn.2.7}) in eqn.(\ref{eqn.2.6}), and separating the integration on $\Delta$ from the derivatives with $x$, we get the equation
\begin{equation}
    f(x,t)+\tau\frac{\partial f}{\partial t} = f(x,t)\int_{-\infty}^{\infty}{\phi(x,\Delta)d\Delta} +\frac{\partial}{\partial x} f(x,t)\int_{-\infty}^{\infty}{\Delta\phi(x,\Delta)d\Delta} + \frac{\partial^2}{\partial x^2}f(x,t)\int_{-\infty}^\infty{\frac{\Delta^2}{2}\phi(x,\Delta)}d\Delta + \cdot\cdot\cdot ,
    \label{eqn.2.8}
\end{equation}
which will generate the diffusion equation, on consideration of the isotropy and normalisation of $\phi(x,\Delta)$ everywhere in space as
\begin{eqnarray}
    \frac{\partial f(x,t)}{\partial t}=\frac{\partial^2}{\partial x^2} D(x)f(x,t),
    \label{eqn.2.9}
\end{eqnarray}
where $D(x)=\int_{-\infty}^{\infty}{\frac{\Delta^2}{2\tau}\phi(x,\Delta)d\Delta}$. 

The diffusion current density corresponding to the above equation is $j_{diff}=-D(x)\frac{\partial f(x,t)}{\partial x}-f(x,t)\frac{\partial D(x)}{\partial x}$, where the second term on the right-hand side in the expression of the current density is the non-Fickian diffusion current which appears on the same footing as the Fickian diffusion current term which is the first one. However, as a standard practice, this non-Fickian diffusion current term is generally neglected when writing the diffusion equation, even in the presence of coordinate-dependence of diffusion, where the diffusion equation is written \cite{crank1979mathematics} as
\begin{equation}
      \frac{\partial f(x,t)}{\partial t}=\frac{\partial}{\partial x} D(x)\frac{\partial}{\partial x}f(x,t).
    \label{eqn.2.10}
\end{equation}
It is not difficult to identify the fact that the eqn.(\ref{eqn.2.10}) is completely inconsistent with Einstein's theory of Brownian motion; however, this equation gets imposed on the basis of the misconception that diffusion is only a concentration inhomogeneity relaxing process. It is indeed strange that people generally consider the resulting diffusion governed by equation (\ref{eqn.2.10}) to be {\it normal diffusion}, even though the equation itself is inconsistent with Einstein's theory generalised to an inhomogeneous diffusion process.

The diffusion equation (\ref{eqn.2.9}) will result in a steady state, equilibrium (when temperature is constant everywhere in space) distribution of particles 
\begin{equation}
    C(x)\equiv f_{eq}(x)=\frac{\alpha}{D(x)},
    \label{eqn.2.11}
\end{equation}
where $\int_{-L}^{L}f(x)dx = N$ is the total conserved number of particles in a confined region, with the equilibrium condition being given by $D(x)f_{eq}(x)=\alpha$, and $\alpha$ is a constant. Note that $1/D(x)$ is the density of state. In the presence of a not-so-strong conservative force $F(x)=-\frac{\partial V(x)}{\partial x}$, the equilibrium solution of the diffusion equation (\ref{eqn.2.9}) will be 
\begin{equation}
    f_{eq}(x)=(\alpha/D(x))e^{-V(x)/k_BT},
    \label{eqn.2.12}
\end{equation}
where the $k_B$ is the Boltzmann constant and the existence of the local Stokes-Einstein relation $D(x)=k_BT/\Gamma(x)$ is taken into account with $\Gamma(x)$ being the local damping coefficient. Note that, in Einstein's theory, the Stokes-Einstein relation is necessarily local since the definition of the diffusivity is necessarily local in this theory. 

It is these equilibrium distributions, as shown in equation (\ref{eqn.2.11}) and (\ref{eqn.2.12}), that apparently defy our existing notion of diffusion as a process that evens out concentration inhomogeneities, whereas we see that diffusion, when inhomogeneous, drives them. This very result comes from a straightforward generalisation of Einstein's theory of Brownian motion without violating any of the postulates on which the theory rests. This can have extremely nontrivial consequences that have not, so far, received adequate attention despite being (correctly, but probably unknowingly) incorporated in the theory of Thermodiffusion.

In the context of the process of Thermodiffusion, Sydney Chapman made this generalisation of Einstein's theory of diffusion on a meticulous basis in kinetic theory in 1928 \cite{chapman1928brownian}. Subsequently, Kiyosi It\^o developed the formal methods of stochastic calculus, in general, for such an inhomogeneous medium \cite{ito1951stochastic}, which also yields the same diffusion equation (\ref{eqn.2.9}). All of these have been largely ignored in the physics literature, driven by the mistaken notion that diffusion is necessarily a transport process that only relaxes concentration inhomogeneities. What is more disturbing is that this restricted notion of diffusion as a transport process is often considered consistent with Einstein's theory of diffusion based on Brownian motion. 

Let us conclude this article by noting that the underlying reason that dictates {\it normal diffusion} cannot be solely a concentration-relaxing process. The reason for this is very general and based on symmetry. The diffusivity (or equivalently, the damping coefficient) does not enter the Hamiltonian of a system. Therefore, independent of the existence of any system in the otherwise homogeneous and isotropic space (in the absence of diffusion), the presence of diffusion characterises the underlying properties of the space in which the Hamiltonian of a system (when present) will introduce additional lifting of symmetries. This is the reason the underlying space retains its homogeneity in the presence of uniform diffusion with constant diffusivity, and it cannot result in anything other than relaxing concentration inhomogeneities, as the homogeneity of space will not allow maintaining any in for otherwise (beyond thermal response) free particles.

However, when diffusion becomes inhomogeneous, for whatever reason, an otherwise free particle experiences an underlying space whose homogeneity and isotropy are broken at large scales at which the diffusion is inhomogeneous. The homogeneity and isotropy are broken at large scales, since, according to Einstein's theory, diffusion is a homogeneous and isotropic transport process at mesoscopic scales. How would an otherwise free particle respond to the presence of these broken symmetries of the underlying space? The particle must respond asymmetrically in accordance with the underlying broken symmetry of space (which is actually due to a hidden thermal/mechanical/hydrodynamical origin), and diffusion in such a situation can drive concentration inhomogeneities. This is a fundamental picture; when kept in mind, it can help us re-examine the long-held notion that {\it normal} diffusion only relaxes concentration inhomogeneities.


%

\end{document}